# Broadband nonlinear optical microresonator array for topological second harmonic generation

Ruoyu Wang, Yiming Pan, Xiaoqin Shen*

School of Physical Science and Technology, ShanghaiTech University, Shanghai, China 201210

Corresponding email: shenxq@shanghaitech.edu.cn

**Abstract**

Topological photonics enables robust light manipulation with third-order optical nonlinearity, yet integrating second-order optical nonlinearity into a topological system faces fundamental challenges: frequency-dependent topological bandgaps impede simultaneous edge states for pump and second harmonic photons at an octave. Here we present a broadband topological nonlinear photonic system via dual frequency topological bandgap engineering in a 2D nonlinear microresonator array. By designing a square lattice with synthetic magnetic fluxes, we achieve topological phase matching—preserving unidirectional edge states for both frequencies while enabling efficient second-harmonic generation. The system exhibits flux-programmable SH chirality, where SH photons reverse propagation direction via Chern number transitions (e.g. $C$ = -1 → +1) without sacrificing robustness. Moreover, we show that the design theoretically yields >100x higher SHG efficiency than single resonators at high powers via topology-enhanced coherent buildup. Our topological SHG works in a parameter regime that can be readily accessed by using existing low-loss integrated photon platforms like thin film lithium niobite.

# Introduction

Topological photonics has emerged as a transformative paradigm for designing robust optical systems[1], leveraging unidirectional, backscattering-resistant edge states analogous to those in electronic topological insulators. These photonic analogs of quantum phenomena[2]—such as the integer quantum Hall effect, quantum spin Hall effect, and non-Hermitian topological states—have unlocked unprecedented control over light propagation and localization. Recent advances have further expanded this field by exploring the interplay between topological physics and optical nonlinearities[3, 4]. Different platforms such as dielectric meta-surfaces[5], and silicon-based waveguide arrays[6] and coupled ring resonator arrays[4, 7] have been demonstrated for third-order ($\chi^{(3)}$) nonlinear processes, including four-wave-mixing, third harmonic generation and parametric amplification. Among these systems, silicon-based two-dimensional (2D) coupled ring resonator arrays exploit strong light confinement and high-quality resonances to achieve enhanced nonlinear efficiencies and robustness against fabrication disorders, enabling the exploration of practical topological photonic devices, such as topological optical delay lines[8], topological lasers[9], and topological single-photon[10] and entangled-photon sources[11]. Recently, 2D coupled ring resonator arrays have also been demonstrated for applications in topological frequency combs[12] and programmable topological photon chips[13].

Despite these advances, the integration of second-order ($\chi^{(2)}$) nonlinearity into topological photonic systems remains a critical challenge. For one reason, typical ring array topological designs fail to preserve edge states for both pump and SH waves that across one octave in frequency due to frequency-dependent bandgap mismatches. For another reason, silicon (the workhorse of integrated photonics) lacks intrinsic second-order nonlinearity, restricting most nonlinear topological studies to third-order effects. One the other hand, however, the pursuit of efficient $\chi^{(2)}$-based devices has driven rapid progress in alternative photonic platforms. With the rapid advances in nanofabrication, emerging material platform such as thin-film lithium niobate[14, 15] and others[16, 17] offer strong $\chi^{(2)}$ nonlinearities and high-quality resonators. These advances position $\chi^{(2)}$ ring resonators as promising building blocks for topological nonlinear optical systems[18]—yet their potential in this context remains unexplored theoretically and experimentally.

Here, we present a broadband topological $\chi^{(2)}$ photonic system via dual frequency topological bandgap engineering in a 2D nonlinear microresonator array. Our system overcomes the fundamental challenge of frequency-dependent topological bandgap misalignment by implementing dual-frequency topological phase matching. By designing a square lattice with synthetic magnetic fluxes, we achieve topological phase matching—preserving unidirectional edge states for both frequencies while enabling efficient second-harmonic generation. The system exhibits flux-programmable SH chirality, where SH photons reverse propagation direction via Chern number transitions (e.g. C=-1 →+1) without sacrificing robustness. From an application perspective, we show that the design theoretically yields >100 times higher SHG efficiency than single resonators at high powers via topology-enhanced coherent buildup. The chiral edge states could enable path-entangled biophotons or quantum entanglement emitters via topology-protected spontaneous parametric down conversion (SPDC) rather than spontaneous four wave mixing. Our design can be readily implemented by using existing low-loss $\chi^{(2)}$ integrated photon platforms like thin film lithium niobite.

## Results and discussion

**The broadband topological system.** Our system consists of an 8 × 8 two-dimensional (2D) square lattice of site ring resonators and link racetrack resonators, coupled via directional couplers (Fig 1a). Photons circulating clockwise (CW) in the site resonators only couple exclusively to counter-clockwise (CCW) photons link resonators. The link resonator length exceeds the site resonators length by $2\eta$, so that the link resonators are off-resonant with the site rings. The linear dynamics of fundamental and second harmonic (SH) photons are both governed by the integer quantum Hall model[19]. A uniform synthetic magnetic field is synthesized via position- and direction-dependent hopping phase between lattice sites, for either fundamental or SH photon (Fig 1b). The tight-binding Hamiltonian describing the linear evolution in the system can be given as

$$\text{H} = \sum_{m,n} \omega_k a_{m,k}^\dagger a_{m,k} - J_k \left( a_{m,k}^\dagger a_{n,k} e^{-i\phi_k} + a_{n,k}^\dagger a_{m,k} e^{i\phi_k} \right), \tag{1}$$

where $k = f, s$ refers the fundamental (*f*) and SH (*s*) fields, and $a_{m,k}^\dagger$ ($a_{m,k}$) are the corresponding photon creation (annihilation) operators at a lattice site $m = (m_x + m_y)$. $\omega_k$ is the frequency of the field $a$ and $\omega_s = 2\omega_f$. $J_k$ is the hopping rate of photons between lattice sites $m$ and $n$. $\phi_k$ is the hopping phase between lattice sites $m$ and $n$. $\phi_k = 4\pi n_{eff,k}(d_{y+1} - d_y)$ gives synthetic magnetic flux

$\phi_k/2\pi$ per plaquette.

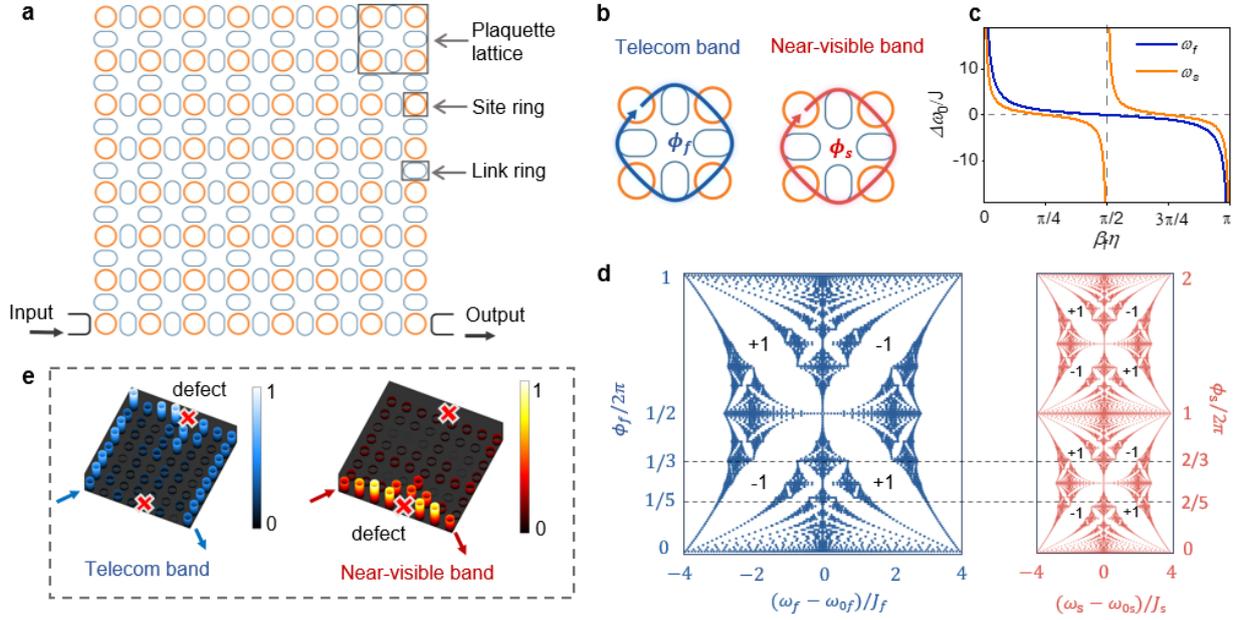

**Figure 1. Schematics of the topological nonlinear photonic system. a.** An 8 × 8 2D square lattice of site ring resonators (orange color) and link racetrack resonators (blue color) coupled to one another through directional couplers. **b.** A single plaquette consisting of four link resonators and four site resonators. Pump photon and SH photon propagating along the plaquette acquire hoping phases $\phi_p$ and $\phi_s$, respectively. **c.** The frequency shift ($\Delta\omega_o$) of the system as a function of $\beta_p\eta$. Divergence of $\Delta\omega_o$ at $\pi/2$ signals the failure of the tight-binding model for SH photon when it is perfectly valid for pump photon. **d.** Hofstadter butterfly spectra. The gaps labelled by the Chern numbers (±1) host back-scattering-immune edge states, whereas bulk bands are topologically trivial (the Chern number is 0). The dashed lines serve as a guide for the eye to show the correspondence of the synthetic magnetic flux and topology between pump and SH photons. **e.** The simulated optical field distributions of the array in telecom and visible frequency regimes. It shows the existence of defect-immune topological edge states for two propagating photons with frequency across an octave.

As shown in Fig 1c, at $\beta_f\eta = (2k + 1)\pi/2$ (k integer, $\beta_f = 2\pi c\omega_f n_{eff,f}$ is the propagation constant of photon at $\omega_f$), the link resonators are perfectly anti-resonant with the site rings for photons at $\omega_f$, leading to an inter-ring hopping phase that validates the tight-binding Hamiltonian at $\omega_f$. However, under this configuration, photons at $\omega_s$ is resonant in link resonators because of $\beta_s\eta = 2k\pi/2$, disrupting inter-ring hopping for photons at $\omega_s$. It leads to the invalidation of the tight-binding Hamiltonian at $\omega_s$. Therefore, equation (1) is valid only when $\beta_f\eta \neq (2k + 1)\pi/2$. In this circumstance, the synthetic magnetic flux $\phi_s/2\pi$ equals to $2\phi_f/2\pi$ when $n_{\text{eff},s} = n_{\text{eff},f}$ (phase-matching condition for SHG), as shown in the Hofstadter spectra (Fig 1d). Bandgaps in the spectra exhibit Chern numbers ($C = \pm 1$), dictating edge-state chirality. Notably, while synthetic magnetic flux

$\phi_f/2\pi$ tuning from 1/5 to 1/3 preserves $C(\omega_f) = -1$, it flips $C(\omega_s)$ from -1 to +1 due to a topological phase transition at $\phi_s/2\pi = 1/2$. It suggests that the broadband topology system could enable nonlinear chirality control: SH photons reverse propagation direction while maintaining topological protection.

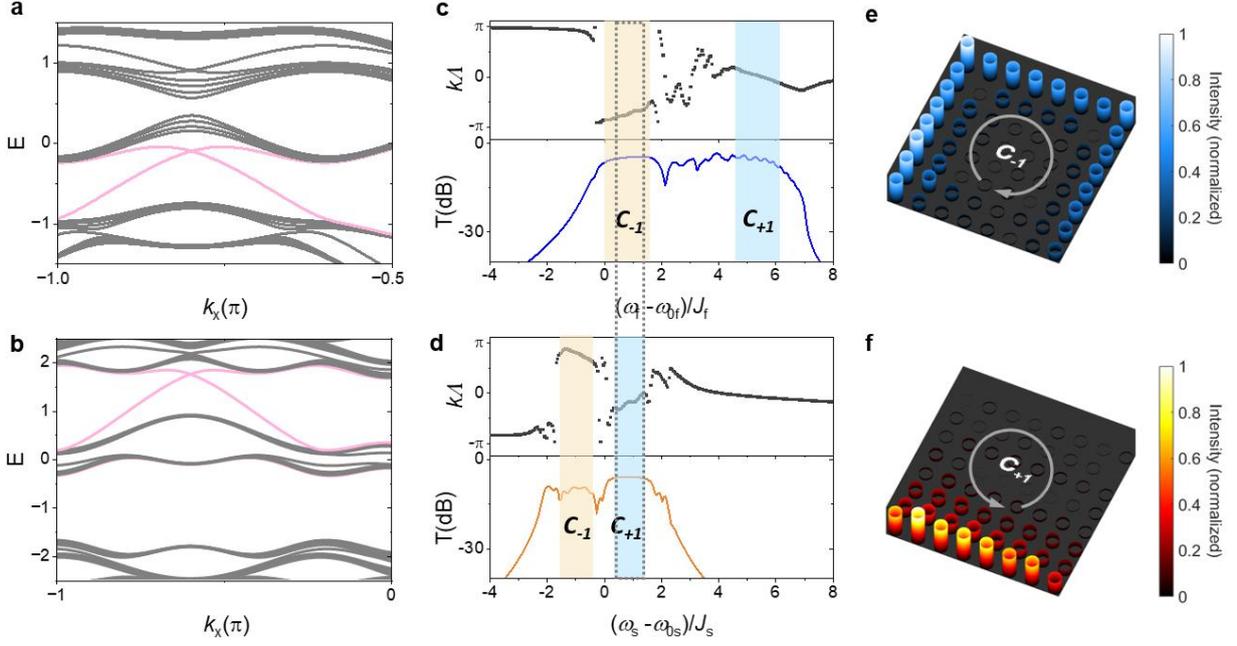

**Figure 2. Representative band diagram and transmission spectra of the system. a-b**, The band diagrams of the system with $\phi_p/2\pi = 0.2$ for photons at $\omega_f$ (**a**) and $\omega_s$ (**b**), respectively. In both cases, the pink color lines indicate that two symmetric edge states form in the gaps. **c-d**, The calculated transmission spectra of the system at $\omega_f$ and $\omega_s$, respectively. The $C_{-1}$ and $C_{+1}$ edge band are highlighted in yellowish and blueish color, respectively. The upper panels in (**c**) and (**d**) show the dispersion relations. $k\Lambda$ is the phase difference between two consecutive resonators at the edge state. **e-f**, The simulated optical field distribution of the system at $\omega_f$ in the $C_{-1}$ edge state (**e**) and at $\omega_s$ in the $C_{+1}$ edge state (**f**), corresponding to regime in the dashed-line box in (**c**) and (**d**), respectively.

**Broadband linear topological behaviors.** For our numerical simulation, we consider an 8 × 8 lattice with $K_{ex}$=2 Ghz, $K_{in}$=0.15 GHz, $J_f$ = 2GHz. We set $\beta_f \eta = \pi/4$, to ensure that the link resonators enable broadband tight-binding Hamiltonian with frequency-dependent hopping phases for photons both at $\omega_f$ and $\omega_s$ (**Fig 1c**). Fig 2a-b shows the calculated energy band diagrams of the 2D array system with $\phi_f/2 = 1/5$ at $\omega_f$ regime (Fig 2a) and $\phi_s/2\pi = 2/5$ at $\omega_s$ regime (Fig 2b). The both two band diagrams feature Dirac-like crossings for the edge states with a pair of symmetric gaps of opposite Chern numbers. **Fig 2c-d** shows simulated transmission spectra of the 2D array for the fundamental resonant mode at $\omega_f$ regime and SH resonant mode at $\omega_s$ regime. The fundamental resonant mode (**Fig 2c**) consists of a $C_{-1}$ edge state band ($C$ = -1) at the relative detuning of ~1, a $C_{+1}$ edge state band ($C = +1$) at the relative detuning of ~ 5 and a bulk state band between the edge bands. The transmission spectrum of the fundamental resonant mode exhibits a shift in center-frequency due to the deviation of the anti-resonant condition of link resonators at $\omega_f$ regime (Fig SX). No center-frequency shift presences for the SH resonant mode (**Fig 2d**) because link resonators are perfectly anti-resonant ($\beta_s \eta = \pi/2$) with the site rings. Importantly, it is found that, at the relative detuning of ~1, the $C_{+1}$ edge state band of the SH mode spectrally aligned with the $C_{-1}$ edge state band of the fundamental resonant mode, as show in the dashed-line box in **Fig2 c-d**. It indicates that dual-resonance could occur in topological protected edge states of the system, which is a prerequisite for efficient SHG.

**Fig 2e-f** shows the simulated optical field distribution of the system by using the transfer matrix method (TMM). Light is coupled in from the left-bottom site ring and is coupled out from the right-bottom site ring. Consistently, it is found that photon at $\omega_f$ propagates clockwise along the short path (the $C_{-1}$ edge state) while photon at $\omega_s$ propagates counter-clockwise along the long path (the $C_+$ edge state). The transmission spectra and optical field distribution of the system at different synthetic magnetic flux $\phi_p/2\pi$ are shown in Fig SX. These results suggest that the 2D array system supports disorder-immune propagation simultaneously for photons at $\omega_f$ and $\omega_s$.

**The SHG nonlinear Hamiltonian.** The generation of SH photon in the system is governed by a three-photon nonlinear interaction described by the Hamiltonian

$$H_{NL} = g \sum_m \left( a^\dagger_{m,f} a^\dagger_{m,f} a_{m,s} + a_{m,f} a_{m,f} a^\dagger_{m,s} \right), \tag{2}$$

where $g$ represents the nonlinear coupling strength and depends on the second-order nonlinear susceptibility ($\chi^{(2)}$) of the material and geometry design of the ring waveguide [20]. Initially, the SH modes reside in the vacuum state as pump photons are introduced into the system. Through the nonlinear interaction, photons are coherently added to or removed from these vacuum modes, resulting in the generation of SH photons. Because of energy conversation, SH photons are generated in longitudinal modes of the site rings at $\omega_s = 2\omega_f$. After the generation, the SH photons evolve under the lattice's linear Hamiltonian $H(\omega_s)$, inheriting robust propagation dynamics.

**Topological SHG behaviors.** To investigate the topological behavior of SHG in the system,

we first simulate the optical field distribution in the array at $\omega_f$, followed by locally generating photons at $\omega_s$ in each site ring accordingly. We then map distribution of the SH field which is governed by:

$$a_s = [I - L(\omega_s)M(2\omega_s)]^{-1}S_{SH}, \qquad (3)$$

where $S_{SH}$ is the SH field vector generated from $H_{NL}$ corresponding to the induced quadratic nonlinear polarization within site resonators. $M(\omega_s)$ describes propagation within individual resonators (including frequency-dependent phases and losses). The diagonal matrix $L(\omega_f) = diag(t_I, \ldots, t_0, 0, \ldots)$ couples the lattice to the bus waveguides, where $t_I$ and $t_0$ represent their transmission rates. Importantly, $S_{SH}$ inherits the topological spatial profile of $a_f$. The subsequent propagation of SH photons is then governed by $M(\omega_s)$, which may exhibit different topology.

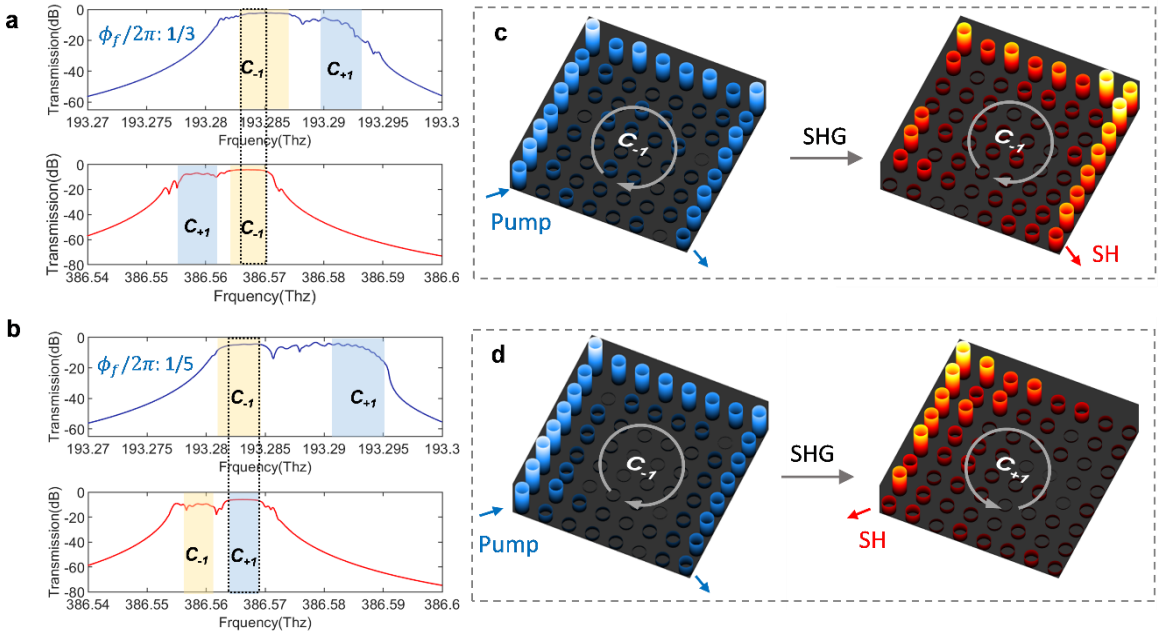

**Figure 3. Representative transmission spectra and optical field distribution of the system. a-b,** Simulated transmission spectra of a fundamental mode (at ~193 THz, upper panel) and an SH mode (at ~386 THz, lower panel) at $\phi_f/2\pi = 1/3$ (**a**) and $\phi_f/2\pi = 1/5$ (**b**). The dashed-line box in each panel highlights the spectral matching band for dual-resonance. **c-d,** Stimulated optical field distribution of the pump and SH in the system corresponding to (**a**) and (**b**). A 1550 nm CW laser at frequency within the dashed-line box is coupled into the system through the left-bottom site ring.

**Fig 3a-b** shows the simulated transmission spectra of a fundamental mode (at ~193 THz) and an SH mode (at ~386 THz) of the system. At $\phi_f/2\pi = 1/3$, the $C_{-1}$ edge state band in the fundamental mode spectrally matches with the $C_{-1}$ edge state band in the SH mode, fulfilling the energy conservation and dual-resonance condition (Fig 3a). Conversely, at $\phi_f/2\pi = 1/5$, the $C_{-1}$

edge state band in the fundamental mode matches with the $C_{+1}$ but not the $C_{-1}$ edge state band the in the SH mode (Fig 3b). It suggests the system exhibits interesting topology-selective dual resonance for SHG

We then inject a continuous wave (CW) pump laser at $\omega_f$ within the edge band of the fundamental mode that enables dual resonance, followed by mapping the optical field distribution of the pump light and the generated SH photon. As shown in Fig 3c, at $\phi_f/2\pi = 1/3$, the generated SH photon propagates clockwise (the $C_{-1}$ edge state) along the longer path of the array, similar to the pump photon. However, at $\phi_f/2\pi = 1/5$ (Fig 3d), it shows that the generated SH photon propagates along the longer edge of the array but counterclockwise (the $C_{+1}$ edge state), while the pump photon propagates clockwise. These results confirm that the SH signal transcribes the |C|=1 state of the pump while its propagation directionality (chirality) is dictated by the band topology at $\omega_s$. In both synthetic magnetic flux configurations, it maintains robust edge propagation due to the system's broadband topological bandgap, which is evidenced by unidirectional trajectories along disorder-prone boundaries. This broadband topology enables SHG with reconfigurable chirality without sacrificing edge protection, which is promising for flux-programmable SH chirality without compromising topological protection could enable a new class of nonlinear isolators.

We further explore the 2D resonator array coupling with one pulley-waveguide to the left-bottom site resonator, as shown in **Fig. 4a**. In this configuration, both pump photon and SH photon are coupled out at a single port. As shown in **Fig 4b**, pump photon propagates along the boundary of the lattice at both $\phi_f/2\pi = 1/5$ and $\phi_f/2\pi = 1/3$, which corresponds to the $C_{-1}$ edge state (clockwise propagating). SH photon also propagates along the boundary of the lattice under the two synthetic magnetic fluxes, no matter in the $C_{-1}$ edge state or $C_{+1}$ edge state.

Finally, we theoretically analyze the SH conversion efficiency of the 2D array, in comparison with that of a single-ring resonator. For a single-ring resonator, the SH conversion efficiency typically reaches its maximum value at a very low critical power, $P^c_{single}$ (μW level), followed by a continuous decline with increasing pump power beyond $P^c_{single}$ due to the pump power depletion or saturation of the intracavity circulating power[17, 21]. It leads to the significant challenge for realizing high conversion efficiency at high pump power or high SH output power[22, 23]. Assuming pump power are evenly distributed in all the site resonators in the edge state, the coherent enhancement from multiple site rings allows the critical power to scale up to the quadratically to the numbers of the site rings (See SI). As shown in **Fig 4c**, in the pulley-waveguide coupled topological system, the simulated result reveal that SH efficiency is more than 100 times higher than that of a single-ring at a pump power of about 20 times of $P_{single}^c$. It can be ascribed to the topology-protected coherent enhancement effect.[24, 25] SH power in both $C_{-1}$ to $C_{-1}$ and $C_{-1}$ to $C_{+1}$ edge state transition starts to surpass that in a single-ring resonator at $P_p \approx 0.3 P_{single}^c$, as shown in **Fig4d**. Profoundly, it can be found that SH power in the system is ideally two-orders of magnitude higher than that in single-ring resonator in high pump power regime ($P_p > 20 P_{single}^c$).

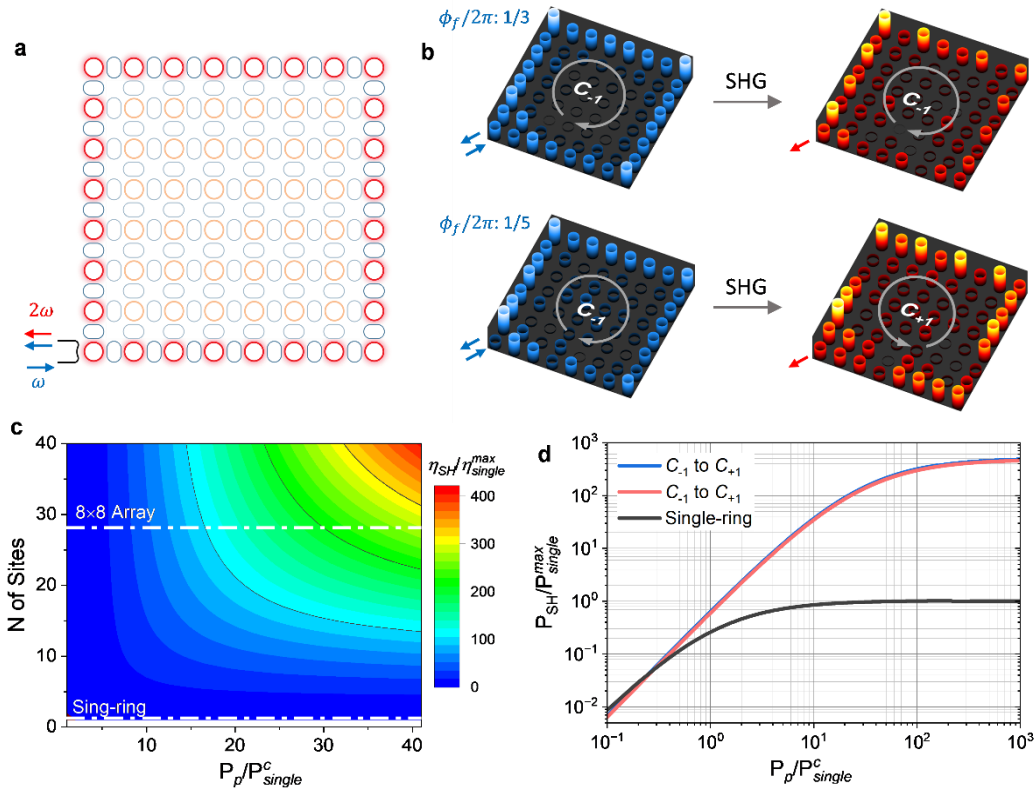

**Figure 4. The pulley-waveguide coupled topological system. a,** Schematic of the system. **b,** Simulated optical field distribution of pump and SH photons in the system at $\phi_f/2\pi = 1/3$ and $\phi_f/2\pi = 1/5$. **c,** SH conversion efficiency $\eta_{SH}$ in a 2D array system as a function of pump power $P_p$ and the numbers of site ring in the edge state. The SH efficiency is normalized to the maximum value $\eta_{single}^{max}$ in a single-ring resonator. The pump power is normalized to the critical pump power $P_{single}^c$ in a single-ring resonator. The dashed lines show the cases of the 8 × 8 array system and a single-ring resonator. **d,** Comparison of SH power $P_{SH}$ in the 8 × 8 array system and in a single-ring resonator as a function of pump power $P_p$. The SH power is normalized to the maximum value $P_{single}^{max}$ in a single-ring resonator.

# Conclusion and outlook

In summary, for the first time, we have established a theoretical framework for broadband topological SHG in a 2D coupled ring microresonator array. By designing a dual-band topological lattice with synthetic magnetic fluxes, we resolved the critical incompatibility between topological protection and $\chi^{(2)}$ nonlinearity across an octave frequency gap. We have demonstrated simultaneous topological protection for pump and SH photons, overcoming bandgap mismatches via flux-dependent Chern number engineering. Theoretical analysis indicates that the topological SHG could have over 100 times higher conversion efficiency than single resonators at high power regime via topology-enhanced coherent buildup. Our work bridges topological photonics and second-order nonlinear optics, demonstrating that $\chi^{(2)}$ processes can inherit topological protection and exhibit flux-programmable SH chirality. The topological design can be readily implemented by using existing low-loss $\chi^{(2)}$ integrated photon platforms like thin film lithium niobate. One could further explore the incorporation of electro-optic tuning in lithium niobate[18], which may enable real-time reconfiguration of synthetic fluxes and programmable topological $\chi^{(2)}$ photonic chip[13] and high speed quantum processors [26].

**Acknowledgements**

This project was supported by the National Natural Science Foundation of China (Grant No. 62275152), Shanghai Pujiang Program (Grant No. 20PJ1411600), and Science and Technology Commission of Shanghai Municipality (Grant No. 20ZR1436400).

**Author contribution**

R. Wang conducted the theoretical work under supervision by X. Shen and fruitful discussion with Y. Pan. X. Shen conceived and designed the project. X. Shen wrote the manuscript with inputs and comments from all authors.

**References:**

1. Ozawa, T., et al., *Topological photonics.* Reviews of Modern Physics, 2019. **91**(1).
2. von Klitzing, K., et al., *40 years of the quantum Hall effect.* Nature Reviews Physics, 2020. **2**(8): p. 397-401.
3. Smirnova, D., et al., *Nonlinear topological photonics.* Applied Physics Reviews, 2020. **7**(2).
4. Mehrabad, M.J., S. Mittal, and M. Hafezi, *Topological photonics: Fundamental concepts, recent*

    *developments, and future directions.* Physical Review A, 2023. **108**(4).
5. You, J.W., Z. Lan, and N.C. Panoiu, *Four-wave mixing of topological edge plasmons in graphene metasurfaces.* Science Advances, 2020. **6**(13): p. eaaz3910.
6. Blanco-Redondo, A., et al., *Topological Optical Waveguiding in Silicon and the Transition between Topological and Trivial Defect States.* Physical Review Letters, 2016. **116**(16): p. 163901.
7. Mittal, S., E.A. Goldschmidt, and M. Hafezi, *A topological source of quantum light.* Nature, 2018. **561**(7724): p. 502-506.
8. Hafezi, M., et al., *Robust optical delay lines with topological protection.* Nature Physics, 2011. **7**(11): p. 907-912.
9. Bandres, M.A., et al., *Topological insulator laser: Experiments.* Science, 2018. **359**(6381).
10. Blanco-Redondo, A., et al., *Topological protection of biphoton states.* Science, 2018. **362**(6414): p. 568-+.
11. Dai, T.X., et al., *Topologically protected quantum entanglement emitters.* Nature Photonics, 2022. **16**(3): p. 248-+.
12. Flower, C.J., et al., *Observation of topological frequency combs.* Science, 2024. **384**(6702): p. 1356-1361.
13. Dai, T.X., et al., *A programmable topological photonic chip.* Nature Materials, 2024. **23**(7).
14. Lu, J., et al., *Periodically poled thin-film lithium niobate microring resonators with a second-harmonic generation efficiency of 250,000%/W.* Optica, 2019. **6**(12).
15. Wang, Z.Y., et al., *Toward ultimate-efficiency frequency conversion in nonlinear optical microresonators.* Science Advances, 2025. **11**(18).
16. Liu, J., et al., *Emerging material platforms for integrated microcavity photonics.* Science China-Physics Mechanics & Astronomy, 2022. **65**(10).
17. Wang, R., et al., *Molecule-induced surface second-order nonlinearity in an inversion-symmetric microcavity.* Optica, 2025. **12**(6): p. 769-773.
18. Hu, Y.W., et al., *Integrated electro-optics on thin-film lithium niobate.* Nature Reviews Physics, 2025. **7**(5): p. 237-254.
19. Mittal, S., E.A. Goldschmidt, and M. Hafezi, *A topological source of quantum light.* Nature, 2018. **561**(7724): p. 502-+.
20. Guo, X., C.-L. Zou, and H.X. Tang, *Second-harmonic generation in aluminum nitride microrings with 2500%/W conversion efficiency.* Optica, 2016. **3**(10).
21. Breunig, I., *Three-wave mixing in whispering gallery resonators.* Laser & Photonics Reviews, 2016. **10**(4): p. 569-587.
22. Guo, X., C.-L. Zou, and H.X. Tang, *Second-harmonic generation in aluminum nitride microrings with 2500%/W conversion efficiency.* Optica, 2016. **3**(10): p. 1126-1131.
23. Lu, J., et al., *Periodically poled thin-film lithium niobate microring resonators with a second-harmonic generation efficiency of 250,000%/W.* Optica, 2019. **6**(12): p. 1455-1460.
24. Harari, G., et al., *Topological insulator laser: Theory.* Science, 2018. **359**(6381).
25. Mittal, S., et al., *Topological frequency combs and nested temporal solitons.* Nature Physics, 2021. **17**(10): p. 1169-1176.
26. Sund, P.I., et al., *High-speed thin-film lithium niobate quantum processor driven by a solid-state quantum emitter.* Science Advances, 2023. **9**(19): p. eadg7268.